\documentclass[a4paper,twoside]{article}

\usepackage{epsfig}
\usepackage{subcaption}
\usepackage{calc}
\usepackage{amssymb}
\usepackage{amstext}
\usepackage{amsmath}
\usepackage{amsthm}
\usepackage{multicol}
\usepackage{pslatex} 
\usepackage{apalike}
\usepackage{algorithm2e}
\usepackage[bottom]{footmisc}
\usepackage{url}
\usepackage[many]{tcolorbox}    	
\usepackage{setspace}               
\usepackage{colortbl}
\usepackage{threeparttable}

\usepackage{SCITEPRESS}     

\definecolor{main}{HTML}{5989cf}    
\definecolor{sub}{HTML}{cde4ff}     

\newtcolorbox{boxE}{
    enhanced, 
    boxrule = 0pt, 
    borderline = {0.75pt}{0pt}{main}, 
    borderline = {0.75pt}{2pt}{sub} 
}

\begin{document} 
\title{Agile Retrospectives: \\
What went well? What didn't go well? What should we do?}
 
  \author{\authorname{  Maria Spichkova\sup{1}, Hina Lee\sup{1}, Kevin Iwan\sup{1}\sup{2}, Madeleine Zwart\sup{1}\sup{2}, Yuwon Yoon\sup{1}, Xiaohan Qin\sup{1} }
  \affiliation{\sup{1}School of Computing Technologies, RMIT University, Melbourne, Australia}   
  \affiliation{\sup{2}Shine Solutions, Melbourne, Australia} 
  \email{maria.spichkova@rmit.edu.au, \{kevin.iwan, madeleine.zwart\}@shinesolutions.com, \{s3910654,s4000117,s3959666\}@student.rmit.edu.au}
  }


\abstract{ 
In Agile/Scrum software development, the idea of retrospective meetings (retros) is one of the core elements of the project process.  
In this paper, we present our work in progress focusing on two aspects: 
analysis of potential usage of generative AI for information interaction within retrospective meetings, and visualisation of retros' information to software development teams. 
We also present our prototype tool RetroAI++, focusing on retros-related functionalities.  \\
~\\
\emph{Preprint. Accepted to the 20th International Conference on Evaluation of Novel Approaches to Software
Engineering (ENASE 2025). Final version to be published by SCITEPRESS, http://www.scitepress.org}}

\onecolumn \maketitle \normalsize \setcounter{footnote}{0} \vfill

\section{\uppercase{Introduction}}
Over the last years, Agile became the most popular approach for software development. This approach gains popularity with each year \cite{al2020agile}. According to the 17th State of Agile report based on the survey conducted in 2023 \cite{Agile17}, 71\% of respondents use Agile in their software development lifecycle, while the most popular Agile methodology continues to be Scrum \cite{schwaber2011scrum}. Moreover, the ideas of Agile are now adopted in various forms in many areas beyond software development. One of the key-elements of the Scrum methodology are \emph{Retrospectives} (\emph{Retros}) - a special type of meetings to be conducted at the very end of each development iteration (\emph{sprint}). The goal of these meetings is to discuss how the sprint went and to identify what could be done to support continuous  improvement within the development team. %
Indeed, at the end of any kind of iteration (whether it is a software development sprint, teaching semester, research project phase or anything else that has any properties of an \emph{iteration}), it makes sense to look back and reflect on it to learn out of the experience. But how exactly do we need to organise this activity?

The idea of retros can really benefit the project only if the participants can have a trusted environment to speak out. In the ideal world, all team members equally respect each other (irrespectively of gender, age, race, etc.), can openly speak about the issues without having a fear to be silenced and lose their face, are happy to suggest ways to improve while knowing that their suggestions will be taken into account. But our real world is not so ideal, and the process of work climate improvement will take years. %
Discussions related to issues, performance and improvements might be very stressful for participants, especially if there is some power imbalance within the team members (e.g., junior vs. senior developers) or some biases might be potentially involved (e.g., related to gender~\cite{marsden2021surfacing}).

One of the solutions would be to use tools to allow for more anonymity in the discussion and to create a psychologically safe environment, see e.g.  \cite{khanna2022your}. Originally,
Scrum retros were conducted as oral discussions with corresponding notes created during the meeting. 
Then (physical) \emph{retro boards} have been introduced, where the space of a white board or a wall was divided in a number of columns or sections, each team member put sticky notes with their comments in the corresponding categories (typically presented by board columns or quadrant), which provided a basis of retro summary and decisions. 
The first two sections are typically representing good and bad points about the sprint, so they are named as \emph{``What went well?''} and  \emph{``What didn't go well?''}, with some wording variations. The rest of the board might be presented different in different approaches. 
 Using an online board provides many advantages, especially in the current software development landscape, where many companies prefer to work in a hybrid mode. Collecting team members' perceptions regarding \emph{``What didn't go well?''} is one of the key drivers to improve team work, therefore it's especially important to make each team member feeling safe while sharing their perceptions on this matter. 
Placing a virtual note on a negative aspects might feel safer than placing a physical sticky note. However, to achieve more anonymity, it should be also hard/impossible to see to which section a person is currently adding a note (e.g., a colleague sitting nearby shouldn't be able to see to which column you currently adding a note). This aspect hasn't been covered yet by the existing tools like Miro, TeamRetro, or Atlassian Retrospective. 

However, a further level of anonymisation is possible:  the inputs on \emph{``What went well?''} and  \emph{``What didn't go well?''} might be collected jointly (i.e., if there is an option  to not manually place a note directly in the corresponding column), and after the collection sorted either manually by Scrum Master, or using sentiment-based automation. For automation, application of Large Language Models (LLMs) might be a promising solution, which is worth to investigate. 

\emph{Contributions:} In this paper, we  
present our preliminary analysis on whether LLMs  might be applicable for this sorting task. We conducted a study on a manually created data set and analysed accuracy of human vs. machine categorisation of the retro comments using OpenAI’s ChatGPT-4 turbo model. We discuss  the lessons learned from this study and the future work that follows from our results.  
We also 
introduce a prototype of a web-based tool \emph{RetroAI++} focusing on its functionality to simply retro-meetings and to make them more safe psychologically.

\section{\uppercase{Related work}}
\label{sec:related}

\subsection{Gamification of Retros}

There are several approaches to enhance Agile/Scrum retros, and many of them propose using gamification, for example,  \cite{MatthiesICSE20,jovanovic2016agile,przybylek2022game,marshburn2018scrum}.   
A case study~\cite{przybylek2017making} has been conducted at Intel Technology Poland, focusing on improving retros  by adopting collaborative games. 
The study~\cite{ng2020playing} presented a replication of~\cite{przybylek2017making}. 
The replication study has been conducted in Bluebay Poland and IHS Markit Gdańsk. The authors also concluded that gamified retros might led to better results than the standard retrospectives. 
Another study conducted at Intel Technology Poland \cite{mich2020retrospective} focused on the use collaborative games.
The results %
confirm the original findings  that game-based retros might improve team members’ creativity, involvement, and communication.  %
While gamification  might provide a good solution to revitalising retros, in our work we mainly focus on aspects related to interaction with data related to project progress and to providing tool-support for releasing potential tensions within retros.  
 
\subsection{Progress overview}
\label{sec:progressOverview} 
Another approach to make the retrospective meetings more efficient, is to provide to the participants an independent overview of their progress. This could be done manually by the Scrum Master or the Product Owner, but manual solution might introduce some biases and lead to additional conflicts. Therefore, it might be useful to get the overview auto-generated. 

\cite{erdougan2018more} analysed how and what kind of historical Scrum project data might be  required for monitoring and  statistical analysis to provide a solid basis for  retrospective meetings, e.g., analysis of the correlation between story
points and actual efforts associated with a product backlog item.  
A resent study conducted by \cite{matthies2021experience} aimed to investigate usage of project data sources into Agile retro meetings, and concluded that a \emph{gather data phase} of might be an important part of a retro meeting. In our prototype, we suggest to go further and to provide the data-based input for the retros as part of the RetroAI++ functionality. 
\cite{gaikwad2019voice} investigated applicability of speech recognition tools, Google Home and Amazon Alexa, for streamlining the retrospective analysis and improving the time boxing of a retrospective by using voice activated commands.
\cite{hakim2024mped} presented a framework for  managing and evaluating changes within Scrum process. The authors didn't focus on providing an input for retro-discussion, however, the elaborated framework might be considered for this purpose.

\subsection{Impact analysis}   
Analysis of teams' satisfaction with retros conducted in their current projects and on issues the teams encounter was presented in the study of \cite{ng2024implementing}. The primary lessons learned of this case study were related to teams' willingness to implement action items and misunderstandings related to the value of discussing positive aspects during
retro meetings. 

A case study conducted in Bosch Engineering GmbH by  \cite{duehr2021positive} led to the conclusion that agile working practices such as retrospectives have a high potential
to improve distributed collaboration. 
The data obtained within this study indicated many aspects of the project work  have been improved after having retros, for example,  
overall quality of the current exchange in the team, 
transparency of information and knowledge in the team, 
 frequency of information and knowledge shared in the team, and  
 variety and reliability of tools used in the team. 
Only one aspect was assessed as being less good after
the retros (compared to before retros), but exactly this aspect is especially alarming. 
While the study of \cite{duehr2021positive} didn't focus on this point deeper, we consider it extremely important: the only negative change was  
\emph{the level of trust in the relationship between team members dropped after the retrospective}.  
This finding might be a critical indication to importance of \emph{how exactly} we conduct the retro meetings. Therefore, our aim is to find solutions that would not solely improve information exchange within the team, but also help creating a a psychologically safe environment and improving the work climate in the team.

A large-scale and cross-sectional survey 
was conducted by \cite{kadenic2023mastering} to investigate the impact of team maturity, team composition, Scrum values, Scrum roles, and Scrum events on the perception of being successful at Scrum. This study established a significant correlation between maturity and the perception of being successful at Scrum. 
There are also a number of studies conducted in the university settings, to analyse students' perceptions of Scrum process, see for example works by  \cite{fernandes2021improving,spichkova2019industry,sun2019software,torchiano2024teaching}.  
In our work, we aim especially on supporting novices, who are especially vulnerable, might be shy to express their thoughts and suggest solutions during the retro meetings. Also, the novices might benefit most from providing providing additional help and more direct, simple instructions on conducting retros. 

\section{\uppercase{Methodology}}
\label{sec:methodology}

 Large Language Models (LLMs) might provide support for completing time consuming and monotonous tasks, where an algorithmic solution doesn't work well. However, the quality of LLM solution might depend on many factors, and one of them is the familiarity of the LLM with the domain language and the context. 
 In this paper, we present our preliminary analysis of applicability OpenAI’s ChatGPT for supporting Agile/Scrum retros.  
 
 In our experiments, we 
 {applied} OpenAI’s GPT-4 Turbo.  
  We created a benchmark dataset $S$ of 200 retro-comments, which we manually annotated using the following four labels:
 \begin{itemize}
     \item 
     \emph{``went well''}: this category included 66  comments (let's denote this set as $S_P$). The set $S_P$ represents 33\% of the benchmark dataset $S$.
     \item
     \emph{``did not go well''}: this category included 99 comments (let's denote this set as $S_N$).  The set $S_N$ represents  49.5\% of the benchmark dataset $S$. This is the largest category because it is typically focus of retro meetings, which goal is continuous improvement.
     \item 
     \emph{``unclear/neutral''}: this category included 28 comments (let's denote this set as $S_U$). The set $S_U$ represents 14\% of the overall set.
     \item 
     \emph{``irrelevant''}: this category included 7 comments (let's denote this set as $S_I$), i.e., 3.5\% of $S$.
 \end{itemize}
 Each comment has been annotated by a single label, multi-labelling has been excluded from our experiment because we consider the specified labels as mutually exclusive. Thus, we have
 \[
 S = S_P ~\cup~ S_N  ~\cup~  S_U ~\cup~  S_I,
 \]
 where 
 $S_P ~\cap~ S_N = \emptyset$, 
 $S_P ~\cap~ S_U = \emptyset$, $S_P ~\cap~ S_I = \emptyset$, 
  $S_N ~\cap~ S_U = \emptyset$, 
 $S_N ~\cap~ S_I = \emptyset$, and 
$S_U ~\cap~ S_I = \emptyset$. 

Based on the dataset $S$, we applied several ChatGPT prompts for auto-grouping/labelling the retro-comments (prompt engineering will be discussed in Section~\ref{sec:prompts}), and analysed the results both quantitatively and qualitatively. 
In our quantitive analysis, we used the following notation: 
\begin{itemize}
    \item $N$ denotes the size of the input set, $N=|S|$. In our experiments, $N=200$. 
    \item 
    $Correct(S)$ denotes the overall set of comments that have been annotated by the LLM correctly, i.e., the set of comments where the category allocation provided by ChatGPT fully matches to the manual annotation. This set consists of four mutually exclusive subsets: 
    $Correct(S_P)$,  $Correct(S_N)$,  $Correct(S_U)$, and $Correct(S_I)$.  
    \item 
    $N_{correct}$  denotes number of comments that have been annotated by the LLM correctly, i.e., $N_{correct} = |Correct (S)|$. 
    \item 
    $Missing(S)$ denotes the overall set of comments that have been provided in the input set, but have been missing in the output set. This is an important indicator of correctness, especially because a situation where some comments disappear might lead to a significant stress by the users (especially in the situation when the users are already under stress due to the nature of the discussion).  
    \item 
    \emph{Nr. of missing comments} ($N_{missing}$) denotes number of comments within the set $Missing(S)$.  
    \item 
    $Dupl(S)$ denotes the overall set of comments that have been allocated by ChatGPT to more than one category at the same time.   
    \item \emph{Nr. of duplicated allocations}  
    denotes number of multi-allocated comments:  $N_{dupl} = |Dupl(S)|$. 
    \item 
    $Incor(S)$ denotes the overall set of comments have been annotated by the LLM incorrectly, i.e., within a wrong category. This set doesn't include the cases of multi-allocations.     
    \item  
    $N_{incor} = |Incor(S)|$ denotes number of comments that have been annotated by the LLM incorrectly. 
    \item 
    \emph{Overall match} ($Match_{overall}$) denotes the percentage of comments correctly annotated by ChatGPT, i.e. comments where the category allocation provided by ChatGPT is the exactly same as manual annotation. For calculation of the overall match, all comments that ChatGPT didn't include in its output are considered as incorrect annotation:
    \[
Match_{overall} = N_{correct} / N
\]
    \item 
    $Match_{simple}$ denotes a simplified representation of the match analysis, where we don't take into account any cases where ChatGPT didn't include comments in its output or reformulated the comments:
    \[
Match_{simple} = N_{correct} / (N-N_{missing})
\]
This metric might be useful for the analysis in the cases when the output set miss a significant number of the items.
\end{itemize}

\noindent
In the current version of RetroAI++ we assume that a comment should be allocated to either \emph{``What went well''} or \emph{``What did not go well''} column. However, it might be possible during a real life retro  
that a user submits a neutral or even an irrelevant comment. It would be unreasonable to allocate such comments to either of above columns,  so we included corresponding categories  in our analysis. 
On the other hand, if a user adds comment to a particular column directly (manually), the content/wording of the comment itself might be more vague/neutral, while the placement in the particular comment will add missing positive/negative context, which we simply cannot have if all comments are places automatically. For example, a comment "Estimation" placed directly in \emph{``What went well''} would mean that someone perceived the effort estimation within the current sprint as good/successful, while the same comment placed directly \emph{``What did not go well''} would mean that someone perceived the effort estimation within the current sprint as inaccurate and requiring improvements. When this comment is submitted for auto-allocation, we cannot know whether it was meant positive or negative without any further context. 
Thus, a better and more practical solution would be to provide both manual and automated options for comment annotation. For these reasons, RetroAI++ provides both functionalities to the users.

\section{\uppercase{Prompt Engineering}} 
\label{sec:prompts}

The aim of prompt engineering is to optimise LLM input to enhance the output performance, see the work of  \cite{white2023prompt}. 
Our engineering strategy was to elaborate instructions as clear as possible by adding more explicit constraints to the input. 
In this paper, due to the space restrictions, we limit our discussion to three prompts to demonstrate the process of elaboration. %
Our overview of quantitive analysis is summarised in Table~\ref{tab:prompts}. 

Prompt 1 has been elaborated to investigate how ChatGPT will work under conditions when the comments should be group in only two categories, i.e., we deliberately excluded categories \emph{``unclear/neutral''} and
     \emph{``irrelevant''}. With this exclusion we aimed to demonstrate the need for having these categories to obtain more precise and meaningful auto-allocation.

\begin{boxE}
\textbf{Prompt 1:} 
A team is doing their Scrum Retrospective and the following comments have been collected. Please group them in two sets ``What went well?" and ``What did not go well". Each comment should be sorted in either ``What went well?" or ``What did not go well": ...
\end{boxE}

\noindent 
Prompt 1 resulted in 48\% of the comments have been missing in the output obtained from ChatGPT. To mitigate this issue, we added the corresponding constraint in the later prompts. 
The overall match value, 
was quite low: only $41\%$. However, this low value was mostly due to many missing comments. If we apply a simplified match analysis, where we don't take into account any missing comments, we obtain   $78\%$,   
which is on a similar level as we obtained for other prompts. 

 Another refinement we applied in Prompts 2 and 3 was specifying a broader set of categories we proposed for manual annotation in Section~\ref{sec:methodology}: Prompt~2  includes   \emph{``unclear/neutral''} but doesn't include
     \emph{``irrelevant''}, while Prompt~3 covers all four categories. 

The overall match value  
was very close by both Prompts 2 and 3, resulting in approx. $74\%$, while $Match_{simple}$ resulted in $77\%$ and $75\%$ with Prompt~2 providing a slightly better match values. The further runs of these prompts resulted in match levels within the same range close to $75-77\%$. 
 Surprisingly, in the executions of Prompt~3,  ChatGPT ignored
the constraints on allocation of the comments to only one category, i.e. we observed allocation of some comments to both categories simultaneously.

      \begin{boxE}
\textbf{Prompt 2:} 
 A team is doing their Scrum Retrospective and the following comments have been collected. Please group them in three sets: ``What went well?", ``What did not go well" and ``Unclear/neutral". Each comment should be sorted in either ``What went well?" or ``What did not go well" or ``Unclear/neutral". Do not reformulate and do not remove any comments. The list of comments: ...
\end{boxE}

\begin{boxE}
\textbf{Prompt 3:} 
 A team is doing their Scrum Retrospective and the following comments have been collected. Please group them in four sets: ``What went well?", ``What did not go well", ``Unclear/neutral" and ``Irrelevant". Each comment should be sorted in either ``What went well?" or ``What did not go well" or ``Unclear/neutral" or ``Irrelevant". Do not reformulate and do not remove any comments. The list of comments: ...
\end{boxE}

\begin{table}[ht!]
    \centering
    \small{ 
    \begin{tabular}{l|l l l}
         & Prompt 1 & Prompt 2 & Prompt 3 \\
         \hline
         Set size & 200 & 200 & 200 \\
         \cellcolor{lightgray!50}$N_{correct}$ & \cellcolor{lightgray!50}81& \cellcolor{lightgray!50}148& \cellcolor{lightgray!50}147\\
         $N_{correct}^P$ & 34& 47&48\\
         $N_{correct}^N$ & 47& 78&77\\
         $N_{correct}^U$ & 0& 23&22\\
         $N_{correct}^I$ & 0& 0&0
         \\
                  \cellcolor{lightgray!50}$N_{incor}$ & \cellcolor{lightgray!50}23& \cellcolor{lightgray!50}45& \cellcolor{lightgray!50}40\\
         $N_{incor}^P$ & 1& 17&17\\
         $N_{incor}^N$ & 16& 17&16\\
         $N_{incor}^U$ & 5& 4&4\\
         $N_{incor}^I$ & 1& 7&3\\
         \cellcolor{lightgray!50}$N_{missing}$ & \cellcolor{lightgray!50}96& \cellcolor{lightgray!50}7 & \cellcolor{lightgray!50}4
         \\
         $N_{missing}^P$ & 31& 2&0\\
         $N_{missing}^N$ & 36& 4&3\\
         $N_{missing}^U$ & 23& 1&1\\
         $N_{missing}^I$ & 6& 0&0\\
         \cellcolor{lightgray!50} $N_{dupl}$ & \cellcolor{lightgray!50}0 & \cellcolor{lightgray!50}0 & \cellcolor{lightgray!50}9 \\
         $N_{dupl}^P$ & 0& 0&1\\
         $N_{dupl}^N$ & 0& 0&3\\
         $N_{dupl}^U$ & 0& 0&4\\
         $N_{dupl}^I$ & 0& 0&1\\
         \cellcolor{lightgray!50}$Match_{simple}$ & \cellcolor{lightgray!50}78\% & \cellcolor{lightgray!50}77\% & \cellcolor{lightgray!50}75\%
         \\
         \cellcolor{lightgray!50}$Match_{overall}$ & \cellcolor{lightgray!50}41\% & \cellcolor{lightgray!50}74\% & \cellcolor{lightgray!50}74\%
    \end{tabular}
    }
    \caption{Quantitive analysis of prompt results}
    \label{tab:prompts}
\end{table}

~\\
It is worth to mention that out of nine comments that have been allocated to multiple (two) categories by ChatGPT, only one comment has been allocated completely incorrectly, while for each of eight other comments one of their  allocations was correct. 
From these observations, we conclude that ChatGPT might struggle with allocation of neutral or irrelevant comments.  
This issue might be mitigated by 
introducing the corresponding rules for conduction retros. 

As Prompt~2 generally provided a slightly better match value, we consider it as a more promising option. 
In the case of Prompt~3, ChatGPT performed not so good mostly because of the issues with multi-allocation of comments, where the majority of the issues were related to having both \emph{unclear/neutral} and \emph{irrelevant} categories.

\section{\uppercase{Lessons learned}} 
\label{sec:results}

In this section we summarise the core lessons learned and briefly discuss solutions we propose to deal with the observed issues while applying ChatGPT for analysis of retrospectives.

\noindent
\textbf{Lesson Learned 1}: 
Even after adding a constraint than none of the comments should be removed or reformulated, some comments have been missing. This happened not at a such large scale as we observed for Prompt~1, but was still significant: while 96 comments have been missed in the response to Prompt~1, 
only 4-7 comments have been missed in case of Prompts~2 and 3 (48\% vs. 2-4\%). This might be a critical issue, as having comments disappeared might create unnecessary stress and tensions. 
\textbf{Proposed solution:} An algorithmic correction in this case might be helpful: We propose to introduce a simple algorithmic check whether all items from the input set $S$ are covered in the auto-allocated sets created by ChatGPT. If some comments have been identified as missing in the auto-allocation, they should be provided to the Scrum Master for manual allocation. As the number of such comments is generally small, the manual allocation  will not be time-consuming.

\textbf{Lesson Learned 2}: 
ChatGPT consistently struggled to categorise  comments that would require knowledge of Agile/Scrum and corresponding terminology, e.g., 
\emph{``Our daily standups were 45 minutes long''}, which is clearly negative from Scrum perspective (meetings of this type should be very short, approx. 10-15 minutes).   
 Another interesting example is \emph{``We played planning poker at the meeting''}: this comment  is clearly positive from Scrum perspective (the team applied a good-practice method for effort estimation), but ChatGPT in different runs labelled it either \emph{irrelevant} or \emph{unclear/neutral} or omitted completely. 
\textbf{Proposed solution:} It would be inefficient to expand a prompt by adding corresponding messages, however, having a pre-trained model might solve the issue. 
Please also note that this identified issue would be irrelevant if the idea of retros is applied outside of Agile/Scrum software development process. 
 
\textbf{Lesson Learned 3}: 
ChatGPT also struggled with vaguely formulated comments and comments including ``but"-statements.   
For example, \emph{``The laptop battery become empty during the demo, but we had a back-up''} is rather positive, because the team resolved their issue successfully. Nevertheless, ChatGPT tends to label it as \emph{unclear/neutral} or omit completely from the output set. 
\textbf{Proposed solution:} It is generally advisable to avoid this type of comments in retros to reduce the cognitive load of other participants. 
A reasonable solution to avoid the issue would be providing to the participants  clear instructions on how the comments should be formulated to facilitate a more productive discussion.

\section{{RetroAI++}}
\label{sec:retroai}

In our RetroAI++ prototype, we aim to automate and refine the practical application of Agile/Scrum processes within Sprint Planning and and Retrospectives. RetroAI++ offers suggestions for sprint organisation as well as  insights for retrospective reflection. 
The prototype combines AI-based planning logic with a more traditional algorithmic foundation in order to enhance the quality of insights produced by the tool. 

\begin{figure}[ht!]
  \centering
  \includegraphics[width=\linewidth]{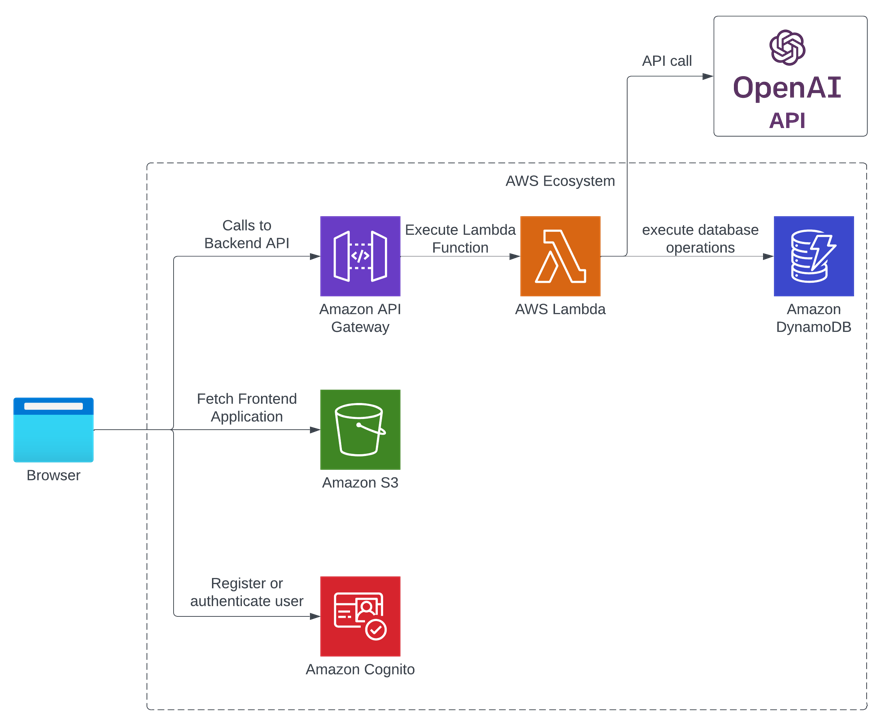}
  \caption{RetroAI++ system architecture}
  \label{fig:RetroAIarch} 
\end{figure}

The general system architecture of our prototype is presented in Figure~\ref{fig:RetroAIarch}. 
The front-end of RetroAI++ has been built using JavaScript and React. For back-end solution, this project uses Java and DynamoDB tables.  
The prototype runs on AWS. 

\begin{figure}[ht!]
  \centering
  \includegraphics[width=0.9\linewidth]{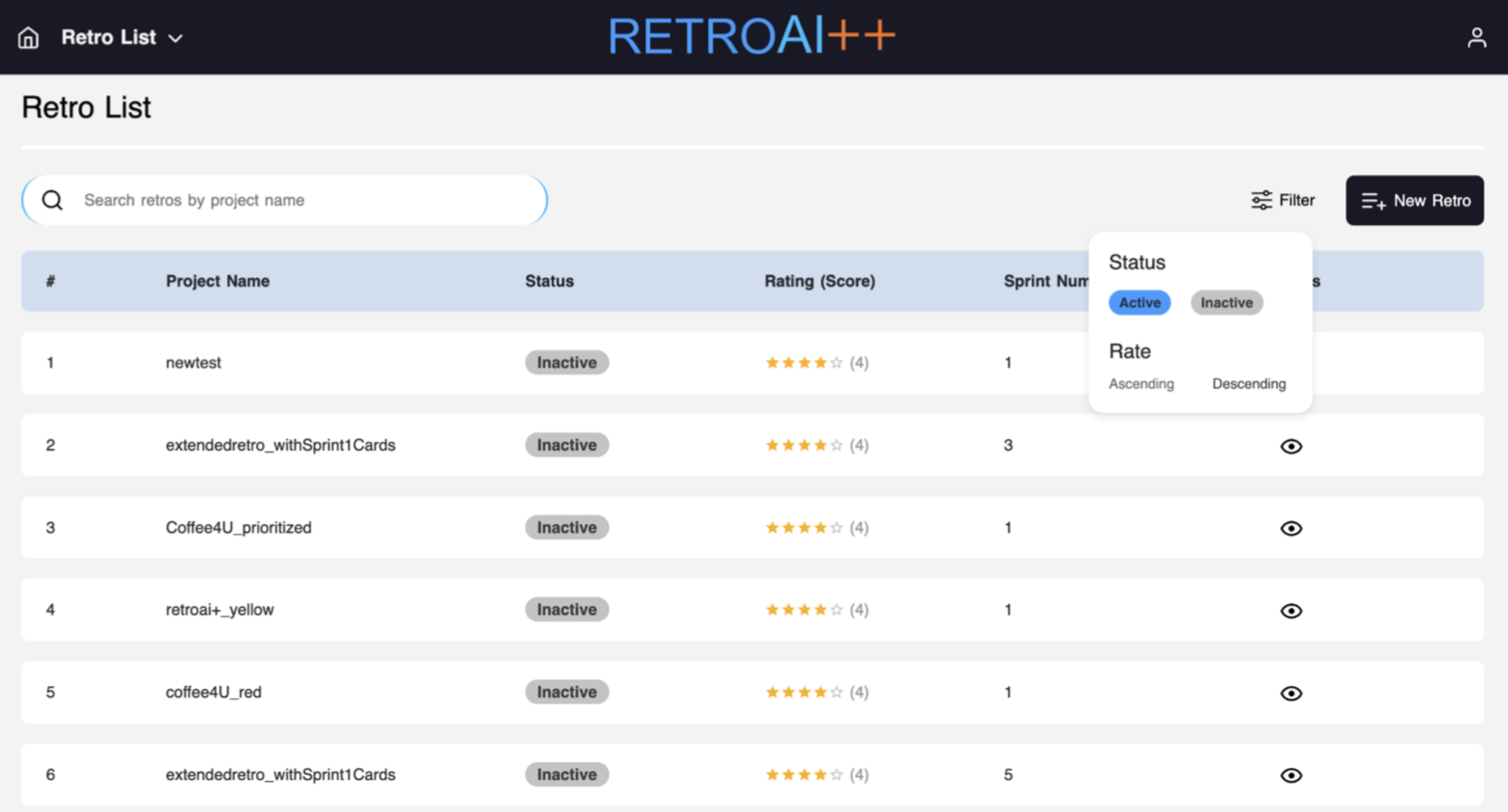}
  \caption{RetroAI++: retro-dashboard}
  \label{fig:RetroAIretros}
\end{figure} 

RetroAI++ provides tool support for sprint planning and retrospective analysis, but in this paper we focus on its functionality dedicated to the facilitation of retrospective meetings (retros) and provide only short overview of other functionalities.  

Figure~\ref{fig:RetroAIretros} presents a retro-dashboard, which
provides the overview of all retro-boards relevant to the user. The retro-dashboard allows to see the following elements useful for the project analysis and retro-meetings:
\begin{itemize}
    \item 
    Names of the projects, for which retro-meetings might be conducted.
    \item 
    The status of each retro-board: \emph{Inactive}  means that the retro-meeting has been completed and users cannot add further comments to the board, while \emph{Active} means that the board is currently active and comments can be added. This functionality might be useful if a team prefers to collect comments before joining together to a meeting. %
    \item 
    Rating/score of retro/meetings, based on participants feedback.
    \item 
    Sprint number, for which the last retro-meeting has been held or is currently in progress.
\end{itemize}
The retro-dashboard also provides search and filtering functionality to simplify finding the relevant project and meeting. This functionality might be especially useful when   Scrum Master, product Owner and/or Scrum team members are involved in multiple project running simultaneously.

In the rest of this section, we would like to discuss in detail three points, which we consider especially important for retro-meetings:
\begin{itemize}
    \item the overall structure of RetroAI++ retro board,
    \item RetroAI++ functionality to group similar comments to decrease cognitive load of users, and
    \item RetroAI++ functionality to present sprint and retro-meeting summary.  
\end{itemize}

\begin{figure}[ht!]
  \centering
  \includegraphics[width=0.9\linewidth]{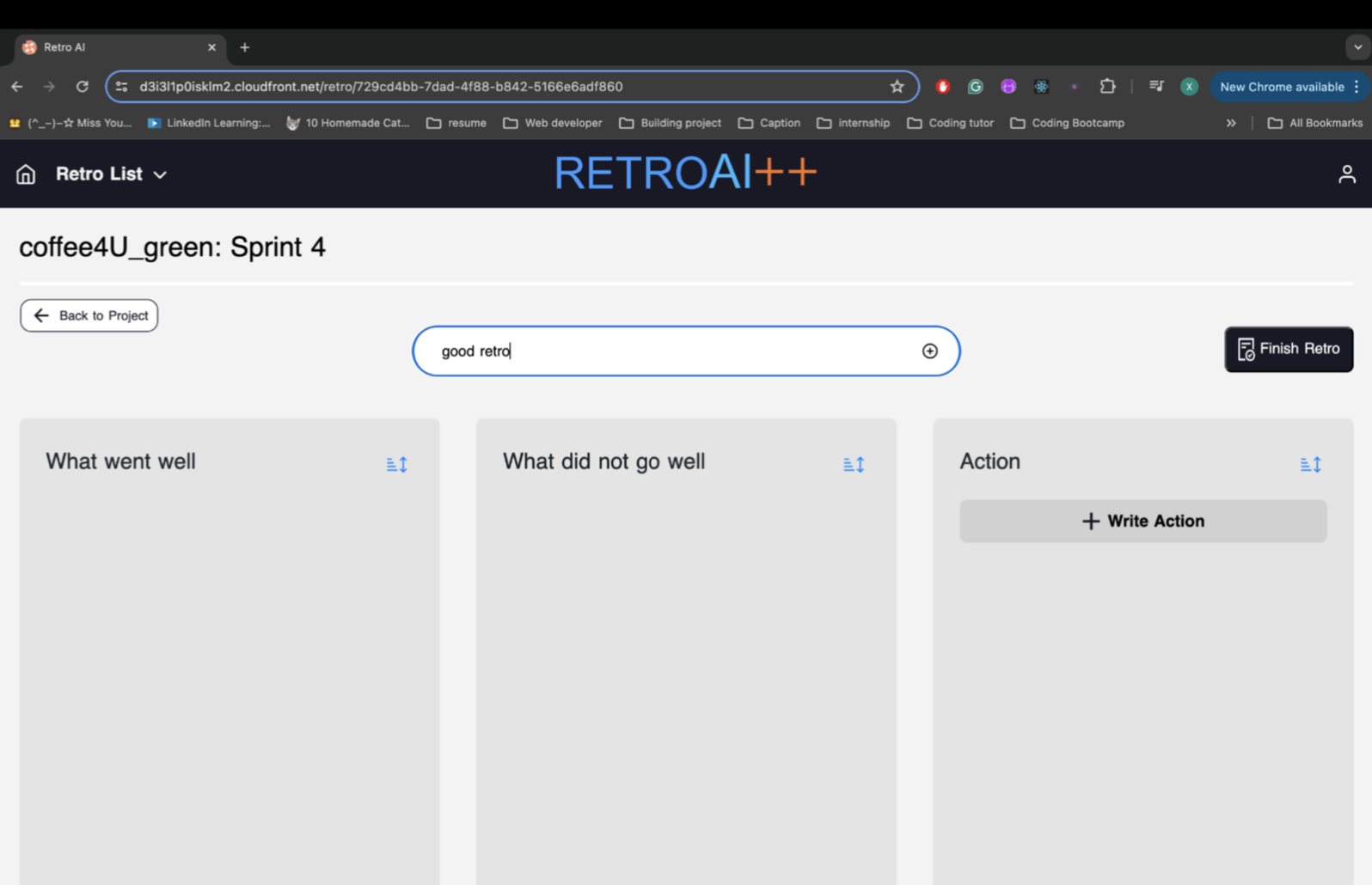}
  \caption{RetroAI++ retro board: General structure (light version of the UI)}
  \label{fig:RetroAI1} 
\end{figure}

Figure~\ref{fig:RetroAI1} presents RetroAI++ retro board, which consists of three columns 
\emph{``What went well''} and  \emph{``What didn't go well''}, and \emph{``Actions''}. 
The advantage of the RetroAI++ retro board is that the comment can be added  in the input field that is located above the columns and isn't associated with any of them, i.e., none can see to which of the columns the input is written. Then the allocation of comments to the columns is done automatically, under support of Open AI.  

An alternative solution would be to get the comments allocated to the columns manually, e.g., by the Scrum Master facilitating the retro meeting, but this would slow down the meeting. In our prototype we use ChatGPT API to provide a preliminary solution for this task.

\begin{figure}[ht!]
  \centering
  \includegraphics[width=0.9\linewidth]{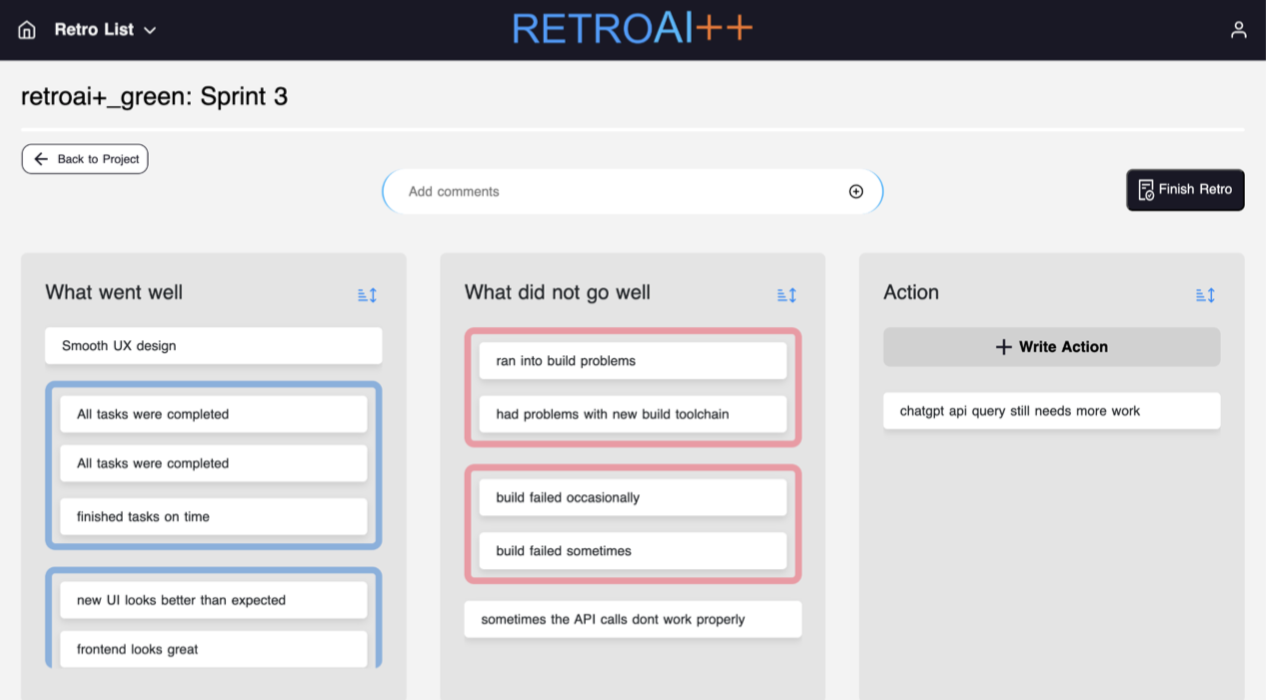}
  \caption{RetroAI++ retro board: grouping similar comments}
  \label{fig:RetroAIgrouping} 
\end{figure}

Often, team members submit comments that are semantically similar: if something was very good or upsetting during a sprint, it's very likely that many team members will have the same feelings about it. It makes sense to visually group similar comments as this decreases the cognitive load of the board analysis. We implemented this idea by presenting similar comments within a group highlighted with a colour frame: blue for \emph{``What went well''}  and red for \emph{``What didn't go well''}, see Figure~\ref{fig:RetroAIgrouping}. 

In the current version of the prototype, grouping has been implemented as a  manual functionality (which can be applied by any team member), but as the future work we would like to explorer application of AI approaches to allow the team have this boring but important task done for them automatically. As preliminary solution, we provide a functionality to sort comments by the frequency of their appearance,  
which might streamline the currently manual process.

 
RetroAI++ can also provide a summary of a sprint, generated using ChatGPT 
based on a Kanban board for this sprint. This information can serve as a starting point for a retro-meeting, as the auto-generated summary provides a short  overview of the team's progress over the sprint wrt. to the Sprint backlog, i.e., wrt. the plan the team had for this sprint.

\section{\uppercase{Threats to validity}}
\label{sec:limitations}

There are several threats to validity of our experiments.  
The first threat is the limited scope of the benchmark dataset, which was limited to 200 retro-comments and created manually, which means it obviously doesn't cover fully the infinite set of all possible retro-comments that can be written in the real life meetings. 
However, our manually created benchmark dataset has a significant size and covers typical points that arise in the retrospectives in real industrial projects. 

The second threat is the limited number of runs for each prompt presented in the experiment analysis in Sections~\ref{sec:prompts} and~\ref{sec:results}. While running our experiments, we observed that the results of prompt executions might differ slightly, i.e., if we run the same prompt multiple times the responses of ChatGPT might be not exactly the same. However, we haven't observed any statistically significant difference, therefore due to space limit we restrict only discussion to the analysis of a single run per each prompt.    
 
Also, the dataset used in our experiment has been created and manually labelled by the second author and then refined and extended by the first author, based on the experience from industrial projects. 
The analysis and classification of the ChatGPT responses has been done 
manually by the authors. To mitigate the issues with incorrect labelling and classification, we used peer-reviewing strategy.

\section{\uppercase{Conclusions}}
\label{sec:conclusions}

In this paper, we presented our ongoing research on streamlining Agile/Scrum processes with the support of AI approaches.  
We  discussed our experiments with OpenAI's ChatGPT-4 turbo to analyse the applicability of generative AI for supporting Agile/Scrum retrospective meetings and summarised the core lessons learned from these experiments. 
We also presented our prototype tool RetroAI++, whose aim is to automate and simplify Agile/Scrum processes for software development projects. We especially focused on RetroAI++ functionality to facilitate retro-meetings.\\
\emph{Future work:} 
As our future work, we plan to conduct  experiments on a larger dataset and to refine/extend our prototype.  

\section*{\uppercase{Acknowledgements}}

We would like to thank Shine Solutions for sponsoring this project under the research grant PRJ00002505, and especially Branko Minic and Adrian Zielonka for priding their industry-based expertise and advices. We also would like to thank students who contributed to creation of earlier versions of the RetroAI tool: 
Weimin Su, Ahilya  Sinha,  
Hibbaan  Nawaz, 
Kartik Kumar,  
Muskan Aggarwal,  
Justin John, Shalvi Tembe, Niyati Gulumkar, Vincent Tso, and Nguyen Duc Minh Tam.

\bibliographystyle{apalike}
{\small

}

\end{document}